\documentclass[aps,amsfonts,pra,twocolumn,showpacs]{revtex4}
\usepackage{epsfig,amsmath,amssymb,bm,epsf,graphics,psfrag,color,ulem}

\def\bra#1{\langle#1\vert}
\def\ket#1{\vert#1\rangle}
\def\ketbra#1{\vert#1\rangle\langle#1\vert}

\def\ipr#1#2{\langle#1\vert#2\rangle}

\def\Longarrow{\protect\@lra}
\def\@lra{\relbar\joinrel\relbar\joinrel\relbar\joinrel%
          \relbar\joinrel\rightarrow}

\begin{document}
\title{Hyperentangled Bell-state analysis}

\author{Tzu-Chieh Wei, Julio T. Barreiro, and Paul G. Kwiat}
\affiliation{Department of Physics, University of Illinois at
Urbana-Champaign, 1110 West Green Street, Urbana, Illinois 61801-3080,
U.S.A.}

\date{\today}

\begin{abstract}
It is known that it is impossible to unambiguously distinguish the
four Bell states encoded in pairs of photon polarizations using only
linear optics. However, hyperentanglement, the simultaneous
entanglement in more than one degree of freedom, has been shown to
assist the complete Bell analysis of the four Bell states (given a
fixed state of the other degrees of freedom). Yet introducing other
degrees of freedom also enlarges the total number of Bell-like states.  We
investigate the limits for unambiguously distinguishing these
Bell-like states. In particular, when the additional degree of
freedom is qubit-like, we find that the optimal one-shot
discrimination schemes are to group the 16 states into 7
distinguishable classes, and that an unambiguous discrimination is
possible with two identical copies.
\end{abstract}
 \maketitle
\noindent {\it Introduction\/.} Just as
the controlled-NOT~\cite{NielsenChuang00} is one of the most important
two-qubit gates in quantum computation, Bell measurement is one of
the most important two-qubit measurements, as it enables many
applications in quantum information processing, such as superdense
coding~\cite{BennettWiesner92,MattleWeinfurterKiwatZeilinger96},
teleportation~\cite{BennettBrassardCrepeauJozsaPeresWootters93,BouwmeesterPanMattleEiblWeinfurterZeilinger97,BoschiBrancaDeMartiniHardyPopescu98},
quantum fingerprinting~\cite{Buhrman01,Horn05}, and direct
characterization of quantum dynamics~\cite{MohseniLidar06}. However,
it was shown that complete Bell-state analysis (BSA) using linear
optics is not
possible~\cite{VaidmanYoran99,LutkenhausCalsamigliaSuominen99}, and
that the optimal probability of success is only
50\%~\cite{LutkenhausCalsamigliaSuominen99,CalsamigliaLutkenhaus00,Calsamiglia02},
for which the optimal BSA schemes have been realized
experimentally~\cite{MattleWeinfurterKiwatZeilinger96,GisinGroup,ZeilingerGroup}.
But Kwiat and Weinfurter (KW)~\cite{KwiatWeinfurter98} showed that
with additional degrees of freedom, such as timing or momentum, it
is indeed possible to achieve complete BSA for  four Bell states,
given that the additional degrees are in a fixed entangled state. Other
similar BSA schemes have also
been proposed~\cite{WalbornPaduaMonken03,WalbornNogueiraPaduaMonken03,RenGuoGuo05}
and implemented~\cite{DeMartini06,Julio07}. In all of these schemes, such
states are called ``hyperentangled''~\cite{Kwiat}, and such measurements are
termed embedded {BSA}~\cite{KwiatWeinfurter98}. Hyperentangled
states with polarization and orbital angular momentum of two photons
have recently been created and characterized~\cite{Julio}.
Furthermore, the KW scheme for BSA has recently been implemented
by Schuck et al.~\cite{SchuckHuberKurtsieferWeinfurter06}.
Nevertheless, adding additional degrees of freedom also enlarges the Hilbert
space, and hence the
number of Bell-like states (e.g. see Table~\ref{tbl:1}); all previous investigations on
embedded BSA have  focused on a subset these states (e.g. states with fixed $\ket{\phi^+}$). It is therefore important to set 
theoretical limits on optimal BSA in the enlarged Hilbert space.

In this Paper, we investigate the optimality of
hyperentanglement-assisted BSA, with both degrees of freedom being
qubit-like, such as polarization ($H$ and $V$) plus {either} two
momenta (spatial directions) or two orbital angular momenta or two time bins. The
resulting Bell-like states for two photons thus total sixteen. We
show that an unambiguous state discrimination is impossible but that
the optimal scheme divides the 16 Bell states into 7 distinct
groups.  We also show by construction that an unambiguous discrimination 
of any of the sixteen states
requires two copies of the same states. Finally, we discuss the
implications for superdense coding, teleportation and quantum
fingerprinting.

\begin{figure}
\vspace{0.5cm} \centerline{ \rotatebox{0}{
        \epsfxsize=7.6cm
        \epsfbox{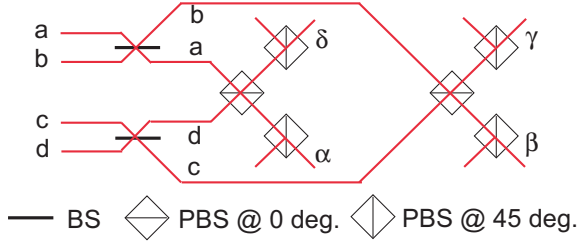}
} }
\caption{ (Color online) Kwiat-Weinfurter scheme
for the embedded Bell-state analysis. } \label{fig:KW}
\end{figure}

\smallskip
\noindent {\it Kwiat-Weinfurter scheme for Bell-state analysis\/.}
KW showed that {when} the momentum degrees of freedom {are} in a
fixed entangled state, the four polarization Bell states can be
unambiguously distinguished~\cite{KwiatWeinfurter98}. Let us introduce 
 the 16 Bell-like states,
constructed from two photons with polarization and momentum (or spatial mode) or timing degrees of freedom: 
(1)
$\{H,V\}\otimes\{a,c\}$ and (2) $\{H,V\}\otimes\{b,d\}$~\cite{basis}. These
states result from the different combinations of the four
polarization Bell states,
\begin{subequations}
\begin{eqnarray}
 \ket{\Phi^\pm}&\equiv &\big(\ket{H}_1\ket{H}_2\pm
\ket{V}_1\ket{V}_2\big)/\sqrt{2}, \\
\ket{\Psi^\pm} &\equiv& \big(\ket{H}_1\ket{V}_2\pm
\ket{V}_1\ket{H}_2\big)/\sqrt{2},
\end{eqnarray}
and the four momentum Bell states,
\begin{eqnarray}
\ket{\phi^\pm} &\equiv &\big(\ket{a}_1\ket{b}_2\pm \ket{c}_1\ket{d}_2\big)/\sqrt{2}\\
 \ket{\psi^\pm} &\equiv &\big(\ket{a}_1 \ket{d}_2 \pm \ket{c}_1 \ket{b}_2\big)/\sqrt{2}.
\end{eqnarray}
\end{subequations}
The detection patterns for the KW scheme (Fig.~\ref{fig:KW})  are shown in
Table~\ref{tbl:1}. The 16 states are divided into 7 distinct
{classes} according to the measurement outcome~\cite{footnote}. Except that one
{class} contains 4 states, all others each have 2 states. Thus, no
single state can be unambiguously distinguished using this scheme.
If the momentum state is $\phi^+$, the four states with
distinct polarization Bell states belong to four distinct classes,
and hence can be distinguished. Similarly, if the polarization
state is $\Phi^+$, the states with four distinct momentum Bell
states can be distinguished. Therefore, the same setup can perform
BSA for either degree of freedom.

\begin{table}[t]
\begin{tabular}{|c|l|l|}
\hline
Class & State & Detector signature \\
\hline1&
 \begin{tabular}{l}
$\Phi^+\otimes\phi^+$,
$\Phi^-\otimes\phi^-$\\
$\Psi^+\otimes\psi^-$, $\Psi^-\otimes\psi^+$ \end{tabular}
 &
 \begin{tabular}{l}
$\alpha_{45}\alpha_{45}$,
$\alpha_{\overline{45}}\alpha_{\overline{45}}$,
$\beta_{45}\beta_{45}$, $\beta_{\overline{45}}\beta_{\overline{45}}$\\
$\delta_{45}\delta_{45}$,
$\delta_{\overline{45}}\delta_{\overline{45}}$,
$\gamma_{45}\gamma_{45}$,
$\gamma_{\overline{45}}\gamma_{\overline{45}}$
\end{tabular}
\\\hline2&
$\Phi^-\otimes\phi^+$, $\Phi^+\otimes\phi^-$ &
$\alpha_{45}\alpha_{\overline{45}}$,
$\beta_{45}\beta_{\overline{45}}$,
$\delta_{45}\delta_{\overline{45}}$,
$\gamma_{45}\gamma_{\overline{45}}$
\\\hline3&

$\Psi^-\otimes\psi^-$, $\Psi^+\otimes\psi^+$ &
$\alpha_{45}\beta_{45}$,
$\alpha_{\overline{45}}\beta_{\overline{45}}$,
$\delta_{45}\gamma_{45}$,
$\delta_{\overline{45}}\gamma_{\overline{45}}$
\\ \hline4&

$\Psi^+\otimes\phi^+$, $\Phi^+\otimes\psi^-$ &
$\alpha_{45}\delta_{45}$,
$\alpha_{\overline{45}}\delta_{\overline{45}}$,
$\beta_{45}\gamma_{45}$,
$\beta_{\overline{45}}\gamma_{\overline{45}}$
\\ \hline5&

$\Psi^+\otimes\phi^-$, $\Phi^-\otimes\psi^-$ &
$\alpha_{45}\delta_{\overline{45}}$,
$\alpha_{\overline{45}}\delta_{45}$,
$\beta_{45}\gamma_{\overline{45}}$,
$\beta_{\overline{45}}\gamma_{45}$
\\ \hline6&

$\Psi^-\otimes\phi^+$, $\Phi^+\otimes\psi^+$ &
$\alpha_{45}\gamma_{45}$,
$\alpha_{\overline{45}}\gamma_{\overline{45}}$,
$\beta_{45}\delta_{45}$,
$\beta_{\overline{45}}\delta_{\overline{45}}$
\\ \hline7&

$\Psi^-\otimes\phi^-$, $\Phi^-\otimes\psi^+$ &
$\alpha_{45}\gamma_{\overline{45}}$,
$\alpha_{\overline{45}}\gamma_{45}$,
$\beta_{45}\delta_{\overline{45}}$,
$\beta_{\overline{45}}\delta_{45}$
\\ \hline\hline *&

$\Psi^\pm\otimes(a_1 c_2-b_1 d_2)$ &
$\alpha_{45}\beta_{\overline{45}}$,
$\alpha_{\overline{45}}\beta_{45}$,
$\delta_{45}\gamma_{\overline{45}}$,
$\delta_{\overline{45}}\gamma_{45}$ \\ \hline

\end{tabular}
\caption{ \label{tbl:1}Detection signature table. $\Phi^\pm\equiv (H_1H_2\pm
V_1V_2)$, $\Psi^\pm \equiv (H_1V_2\pm V_1H_2)$, $\phi^\pm \equiv
(a_1b_2\pm c_1d_2)$, and $\psi^\pm \equiv (a_1 d_2 \pm c_1 b_2)$.
The subscript $45$ indicates the port associated with transmission
through the
polarizing beam splitter and $\overline{45}$  that
with reflection. The final row lists a unique detection signature, corresponding, however, to states outside the Hilbert space spanned by the 16 hyperentangled Bell states~\cite{footnote}.}

\end{table}

\smallskip
\noindent {\it Optimal hyperentangled Bell-state analysis\/.}
One may wonder what  the optimal Bell-state analysis is.
Calsamiglia~\cite{Calsamiglia02} showed that any element $\ketbra{u_i}$ in a generalized
measurement (i.e., POVM $\sum_i\lambda_i\ketbra{u_i}=\openone$, with $\sum_i\lambda_i=1$)  on
two i-qudits (qudits composed of identical particles) of linear optics can have a Schmidt number
 at most of 2.  As our hyperentangled Bell states have Schmidt number 4,
this means that no single state can be
distinguished from any other, and so unambiguous and complete
BSA for the 16 states is not possible. Thus, the optimal scheme
groups the states into classes, in our case, at most 8 distinguishable
classes.  However, our  analysis of the KW scheme (Table~\ref{tbl:1}) identifies only 7
classes. Now we shall prove that  7 is in fact the upper limit.

We utilize the method of van Loock and
L\"utkenhaus to test whether {8 classes can
be discriminated}.  They showed that a necessary condition for the
distinguishability of {the }states $\psi_i$ and $\psi_j$ ($i\ne j$)
is~\cite{LoockLutkenhaus04}
\begin{equation}
\label{eqn:criterion}
 \langle \psi_i|c_s^\dagger
c_s|\psi_j\rangle=0 \qquad\text{with} \qquad c_s=\sum_{i=1}^N\,\nu_i
c_i,
\end{equation}
where $c_s$ is the annihilation operator, 
linearly composed of $N$ modes (both input and auxiliary) via some
unitary transformation,
and thus the $\nu_i$'s cannot all be zero.  The rationale behind
Eq.~(\ref{eqn:criterion}) is that in order for $\psi_i$ and $\psi_j$
to be distinguishable, the remaining states  should maintain orthogonality
after a single-photon
detection at mode $s$.  In addition,
ancillary photons do not assist state discrimination if either input
or auxiliary states have a fixed number of photons. This means that,
in Eq.~(\ref{eqn:criterion}),  $N$ can be set {as} the number of
input modes.

For the setup shown in Fig.~\ref{fig:KW}, we relabel the input modes
as $\ket{1}\equiv\ket{H}\otimes\ket{a},
\ket{2}\equiv\ket{H}\otimes\ket{c},
  \ket{3}\equiv\ket{V}\otimes\ket{a}, \ket{4}\equiv\ket{V}\otimes\ket{c},
  \ket{5}\equiv\ket{H}\otimes\ket{b}, \ket{6}\equiv\ket{H}\otimes\ket{d},
  \ket{7}\equiv\ket{V}\otimes\ket{b}$ and
  $\ket{8}\equiv\ket{V}\otimes\ket{d}$, where $H$ and $V$ denote the
  polarization degree of freedom and $a$, $b$, $c$ and $d$ denote the
  momentum or direction (or angular-momentum) degree of freedom.  {Thus,
  the} Bell states can be written {as}
\begin{equation}
\label{eqn:Bell}
 \ket{\Psi^{(\mu)}}=\sum_{i,j=1\dots8} W^{(\mu)}_{ij}
c_i^\dagger c_j^\dagger\ket{0},
\end{equation}
where the symmetric matrices $W^{(\mu)}$ are $8\times8$ invertible
(i.e., with nonzero determinant) and characterize the sixteen
($\mu=1\dots 16$) Bell states. If the optimal BSA groups the 16 Bell
states into 8 classes, there must exist  sets of 8 states for which the
conditions set by Eq.~(\ref{eqn:criterion}) are satisfied. On the
other hand, if 7 is the optimal number of classes, no set of 8
states satisfy Eq.~(\ref{eqn:criterion}). To see whether the former
or the latter is true, we have to check whether
Eq.~(\ref{eqn:criterion}) can be satisfied for all possible
combinations of 8 out of the 16 Bell states ($C^{16}_8=12870$, though this  number can be reduced by considering the group
structure of operations that transform the 16 states onto
themselves.)

First, as an example, take two states from
{class} 1 and one from each of the other 6 {classes}:
$\Phi^+\otimes\phi^+,\Phi^-\otimes\phi^-,\Phi^-\otimes\phi^+,\Psi^-\otimes\psi^-,
\Psi^+\otimes\phi^+,\Psi^+\otimes\phi^-,\Psi^-\otimes\phi^+$, and
$\Psi^-\otimes\phi^-$.
Applying Eq.~(\ref{eqn:criterion}) to these states, we have, after
simplifying the equations,
\begin{subequations}
\begin{eqnarray}
&&|\nu_1|=|\nu_3|, |\nu_2|=|\nu_4|, |\nu_5|=|\nu_7|, |\nu_6|=|\nu_8| \\
&&|\nu_1|^2+|\nu_5|^2=|\nu_2|^2+|\nu_6|^2 \\
&&\nu_7^*\nu_5=\nu_2^*\nu_4=\nu_6^*\nu_8=\nu_3^*\nu_1=0.
\end{eqnarray}
\end{subequations}
These lead to the only solution $\nu_i=0$, which is a contradiction.
This shows that one cannot discriminate any state from the above
eight states.

We check all 12870 cases by programming {\tt MATHEMATICA} to examine
the conditions derived from Eq.~(\ref{eqn:criterion}), supplemented
by the normalization condition $\sum_i |\nu_i|^2=1$. This is
achieved by first enumerating and simplifying equations generated
from Eq.~(\ref{eqn:criterion}), as well as the normalization
condition, and then by using the function {\tt FindInstance[\,\,]}
to find an instance of solutions. One feature of {\tt
FindInstance[\,\,]} is that it will {\it always\/} find a solution if there
is one. For all the 12870 cases, {\tt FindInstance[\,\,]} returns an
empty set, showing no solution. Therefore, we conclude that it is
impossible to reliably distinguish among any set of 8 Bell-like
hyperentangled states, and that 7 is the optimal, as is realized in
the KW scheme.

\begin{figure}
\centerline{
\rotatebox{0}{
        \epsfxsize=5.8cm
        \epsfbox{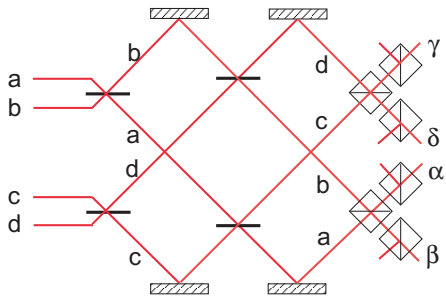}
} }
 \caption{(Color online) Modified
KW scheme. } \label{fig:KW2}
\end{figure}

\begin{table}[t]
\begin{tabular}{|c|l|l|}
\hline
Class& State & Detector signature \\
\hline 1' &
 \begin{tabular}{l}
$\Phi^+\otimes\phi^-$,
$\Psi^-\otimes\phi^-$\\
$\Phi^+\otimes\psi^+$, $\Psi^-\otimes\psi^+$ \end{tabular}
 &
 \begin{tabular}{l}
$\alpha_{45}\alpha_{45}$,
$\alpha_{\overline{45}}\alpha_{\overline{45}}$,
$\beta_{45}\beta_{45}$, $\beta_{\overline{45}}\beta_{\overline{45}}$, \\
$\delta_{45}\delta_{45}$,
$\delta_{\overline{45}}\delta_{\overline{45}}$,
$\gamma_{45}\gamma_{45}$,
$\gamma_{\overline{45}}\gamma_{\overline{45}}$
\end{tabular}
\\ \hline 2' &
$\Phi^-\otimes\phi^-$, $\Phi^-\otimes\psi^+$ &
$\alpha_{45}\alpha_{\overline{45}}$,
$\beta_{45}\beta_{\overline{45}}$,
$\delta_{45}\delta_{\overline{45}}$,
$\gamma_{45}\gamma_{\overline{45}}$ \\
\hline 3' &

$\Psi^+\otimes\phi^-$, $\Psi^+\otimes\psi^+$ &
$\alpha_{45}\beta_{45}$,
$\alpha_{\overline{45}}\beta_{\overline{45}}$,
$\delta_{45}\gamma_{45}$,
$\delta_{\overline{45}}\gamma_{\overline{45}}$ \\ \hline 4' &
$\Psi^+\otimes\phi^+$, $\Phi^-\otimes\psi^-$ &
$\alpha_{45}\delta_{45}$,
$\alpha_{\overline{45}}\delta_{\overline{45}}$,
$\beta_{45}\gamma_{45}$,
$\beta_{\overline{45}}\gamma_{\overline{45}}$ \\ \hline 5' &

$\Phi^+\otimes\psi^-$, $\Psi^-\otimes\psi^-$ &
$\alpha_{45}\delta_{\overline{45}}$,
$\alpha_{\overline{45}}\delta_{45}$,
$\beta_{45}\gamma_{\overline{45}}$,
$\beta_{\overline{45}}\gamma_{45}$ \\ \hline 6' &

$\Phi^+\otimes\phi^+$, $\Psi^-\otimes\phi^+$ &
$\alpha_{45}\gamma_{45}$,
$\alpha_{\overline{45}}\gamma_{\overline{45}}$,
$\beta_{45}\delta_{45}$,
$\beta_{\overline{45}}\delta_{\overline{45}}$ \\ \hline 7'&

$\Phi^-\otimes\phi^+$, $\Psi^+\otimes\psi^-$ &
$\alpha_{45}\gamma_{\overline{45}}$,
$\alpha_{\overline{45}}\gamma_{45}$,
$\beta_{45}\delta_{\overline{45}}$,
$\beta_{\overline{45}}\delta_{45}$ \\ \hline

\end{tabular}
\caption{ \label{tbl:2}Detection signature for the scheme
in Fig.~\ref{fig:KW2}. }

\end{table}

\smallskip
\noindent {\it Unambiguous Bell-state discrimination\/}. Having
seen that a one-shot measurement is unable to perfectly discriminate any
Bell state, it seems natural to ask how many copies are necessary to
enable such discrimination. We show here by construction that 2 copies are sufficient. First, we introduce a slightly modified
measurement scheme from that of KW, shown in Fig.~\ref{fig:KW2}. The
corresponding detection patterns are shown in Table~\ref{tbl:2}.
From Tables~\ref{tbl:1} and~\ref{tbl:2} we see that no two states
share the same class of detector signature. Therefore, we imagine letting one
copy go through the KW scheme and the other through the scheme in
Fig.~\ref{fig:KW2}. Suppose we obtain signatures in 1 and 2'.
Combining both outcomes enables us to uniquely determine which of the
16 states was analyzed, e.g.,  $\Phi^-\otimes\phi^-$ in the example
given~\cite{Unambiguous}.

\noindent {\it More degrees of freedom\/.} We have shown that with
one additional qubit-like degree of freedom for each photon, there exist 7
states (out of 16) that can be distinguished from one another. Next
we consider for each photon $n$ qubit-like degrees of freedom in total.  
 In this case there are
$4^n$ Bell-like states. What is the maximum number of distinguishable subsets  of these states?

Let us begin by noting that we can express the $4^n$ Bell-like
states in the form of Eq.~(\ref{eqn:Bell}), where the upper limit in
the sum is now the number of input modes, $2^{n+1}$. The matrices
$W^{(\mu)}$ are now $(2^{n+1})\times(2^{n+1})$. If one make{s} a
unitary transformation of the modes (using the fact that one can
take the number of modes equal to the number of input modes, ignoring
any auxiliary mode), $a_i^\dagger= \sum_j U_{ij} c_j^\dagger$, the
necessary condition for discrimination between states $\Psi^{(\mu)}$
and $\Psi^{(\nu)}$ ($\mu\neq \nu$) is
\begin{equation}
\bra{ \Psi^{(\mu)}}a_i^\dagger a_i \ket{\Psi^{(\nu)}}=0
\Leftrightarrow \ipr{ \psi^{(\mu)}_i}{\psi^{(\nu)}_i}=0,
\end{equation}
where we have defined $\ket{ \psi^{(\mu)}_i}\equiv a_i
\ket{\Psi^{(\nu)}}$. Because of the unitarity of $W$ and $U$,
$\ket{\psi^{(\mu)}_i}$ has nonzero norm and is equivalent to a
$2^{n\!+\!1}$-component vector. The above orthogonality condition then
implies that there can be at most $2^{n\!+\!1}$ linearly-independent
vectors of $\psi^{(\mu)}_i$ for fixed $i$. Thus, we see that the
maximum number of Bell states that can be distinguished is bounded
above by $2^{n\!+\!1}$. This means that the ratio of the maximal number
of mutually distinguishable sets of Bell states to the total
number of Bell states decreases exponentially with $n$:
$2^{n\!+\!1}/4^n=2^{-n\!+\!1}$.

 We conjecture that
$2^{n+1}-1$ is a good upper bound, as it is true for $n=1$ (e.g.,
polarization only) and $n=2$ (e.g., polarization plus two spatial
modes). Generalizing to different dimensions  of the
degrees of freedom, the absolute upper bound on distinguishable Bell states
can be shown to be $2 d_1d_2d_3\cdots
d_n$.

\smallskip
\noindent {\it Implications  for quantum
communication\/}.
{a) \it Superdense coding\/}. Given that we can choose 7 Bell states
such that they can be distinguished from one another, we can then
take one of them as a shared entanglement and use 7 operations,
taking the state to itself or 6 others, to encode 7 messages. For
example, Alice and Bob share $\Psi^-\otimes\psi^-$. She can locally
transform the state into 6 other states, $\Phi^+\otimes\phi^+,
\Phi^-\otimes \phi^+, \Psi^+\otimes\phi^+, \Psi^-\otimes\phi^+,
\Phi^-\otimes\psi^-,$ and $\Phi^-\otimes\psi^+$. As these seven
states can be distinguished using the KW scheme, Bob can uniquely  determine the message encoded  by Alice, giving a
superdense coding of $\log_2 7\approx 2.8$ bits. For two photons
entangled only in polarization, a superdense coding  encodes only
$\log_2 3\approx 1.58$ bits~\cite{MattleWeinfurterKiwatZeilinger96}.
Even though its extension to two pairs encodes $2\log_2 3\approx
3.17$ bits, the four-photon detection efficiency $\eta^4$ is
typically much smaller than the two-photon efficiency $\eta^2$,
where $\eta$ is the single-photon detection efficiency (usually much
smaller than $70\%$). In fact, as long as the efficiency is less than
$\sqrt{7/9}\approx 88\%$, the single-pair hyperentangled scheme is
superior. Thus, hyperentanglement for superdense coding seems
 more practical than multi-pair entanglement.

{b) \it Quantum fingerprinting\/.} Fingerprinting is a communication
protocol in which two parties, Alice and Bob, want to test whether
they receive the same message from a supplier, but as they cannot
have direct communication with each other. Therefore, they have to communicate
{through a} third party to test whether the two messages are the
same.  Instead of sending the whole messages, they send the
corresponding ``fingerprint'' (a much shorter message) of their
messages to the third party.  {A quantum} protocol is superior to
its classical counterpart {because the former allows } 100\%
{fingerprinting success}. It was shown that shared two-qubit Bell
states enable perfect fingerprinting of {binary-encoded $\{0,1\}$}
messages~\cite{Buhrman01,Horn05}. Here, we propose using
hyperentanglement of a pair of photons to achieve perfect
fingerprinting of $\{0,1,\dots,6\}$ encoded messages. Analogously to
dense coding with hyperentanglement, Alice and Bob share the state
$\Psi^-\otimes\psi^-$, and both parties can locally transform the
shared state into the 7 states: $\Psi^-\otimes\psi^-$, $\Phi^+\otimes\phi^+, \Phi^-\otimes
\phi^+, \Psi^+\otimes\phi^+, \Psi^-\otimes\phi^+,
\Phi^-\otimes\psi^-,$ and $\Phi^-\otimes\psi^+$.  {Thus, they encode
their fingerprints locally by applying the required operations, and
a referee can perform the BSA on the resulting two-photon state to
determine whether  the  fingerprints are the same.

{c) \it Quantum teleportation\/}. A shared Bell-like state enables
{the} teleportation of an unknown state. However, as complete
{BSA} of a two-photon polarization {state} alone is not possible,
schemes employing additional degrees of freedom have been
proposed~\cite{WalbornPaduaMonken03,KwiatWeinfurter98}. The embedded
Bell-analysis schemes proposed in
Refs.~\cite{WalbornPaduaMonken03,RenGuoGuo05,DeMartini06}, however,
cannot be used for teleportation, as their measurements do not
require two photons to interfere, and can be performed locally.  If
these schemes could enable teleportation, it would imply that
entanglement can be created locally by distant parties; but it is
well known that local operations and classical communication cannot
generate entanglement. {Our analysis shows that the KW scheme
enables the teleportation of an arbitrary state encoded in either 
polarization or momentum (not both) with a 50\% probability of
success, the same probability as the two-photon polarization BSA.}
{Suppose} a photon in Alice's laboratory is in a state with known momentum
but arbitrary polarization,$
\ket{\psi}=\big(\alpha\ket{H}_1+\beta\ket{V}_1\big)\otimes\ket{h}_1$,
where $\{h,v\}$ is used to indicate its momentum degree of freedom.
Alice and Bob share the Bell state $(\Phi^+\otimes\phi^+)_{23}$ of
photons 2 and 3. If Alice performs the KW BSA on
photons 1 and 2, there is a 50\% probability (and she knows whether
it succeeds) that Bob can transform his photon into {the} state
$(\alpha\ket{H}_1+\beta\ket{V}_1)$ by performing the corresponding
local operation according to Alice's measurement outcome, and
post-selecting the photon from his momentum modes $b$ or $d$ in
$\phi^+=(a_1b_2+c_1d_2)$.  Similarly, an arbitrary momentum state
$\ket{H}\otimes (\alpha\ket{h}+\beta\ket{v})$ can be teleported. The
use of hyperentanglement of photons, unfortunately, does not {offer}
advantages for teleportation over the conventional polarization-only
teleportation~\cite{BouwmeesterPanMattleEiblWeinfurterZeilinger97,BoschiBrancaDeMartiniHardyPopescu98},
both having only 50\% probability of success.

\smallskip
\noindent {\it Concluding remarks\/.} We have investigated the
optimal Bell-state analysis using projective measurements in linear
optics for hyperentangled Bell states. The results are relevant as
there has been recent experimental progress in realizing BSA of
hyperentangled
states~\cite{DeMartini06,SchuckHuberKurtsieferWeinfurter06,Julio07}. In
particular, we have shown that when the additional degrees of
freedom {are also} qubit-like, the resulting 16 Bell-like states can be, at
best, divided into 7 distinct {classes}. Moreover, we have provided
a method to unambiguously discriminate any of the 16 Bell
states, given two copies of the state.  We have also discussed the
implications for superdense coding, fingerprinting and
teleportation.  We conclude with two open issues for future
study: 1) how generalized measurements might be used to help Bell analysis
in general; and 2) whether other methods such as that of
Eisert~\cite{Eisert05} may provide alternative approaches to understand the
results presented here.

\noindent {\it Acknowledgments\/}. The authors acknowledge useful
discussions with R.
 Rangarajan and N. L\"utkenhaus. This work was supported by  DOE DEFG02-91ER45439, NSF EIA01-21568,
 DTO DAAD19-03-1-0282 and  MURI Center for Photonic Quantum Information Systems.


\begin{thebibliography}{99}
\bibitem{footnote} There is one  
class of four detector outcomes missing. However, this class, indicated
in the last line of Table~\ref{tbl:1}, could be
realized, e.g., by the following two states:
$(H_1V_2-V_1H_2)(a_1c_2-b_1d_2)$ and
$(H_1V_2+V_1H_2)(a_1c_2-b_1d_2)$, which reside outside the Hilbert
space spanned by the 16 Bell states {and are} composed of photon 1
having spatial modes $a$ and $b$ and photon 2 having $c$ and $d$.
\end{thebibliography}
\end{document}